%% file: main.tex
\documentclass[sigconf,nonacm]{acmart}

\AtBeginDocument{%
  \providecommand\BibTeX{{%
    \normalfont B\kern-0.5em{\scshape i\kern-0.25em b}\kern-0.8em\TeX}}}

\setcopyright{acmcopyright}
\copyrightyear{2018}
\acmYear{2018}
\acmDOI{XXXXXXX.XXXXXXX}

\usepackage{booktabs}
\usepackage{mathtools}
\usepackage{amsmath}
\usepackage{multirow} 
\usepackage{makecell}
\usepackage{times}
\usepackage{epsfig}
\usepackage{fancyhdr}
\usepackage[normalem]{ulem}
\usepackage{soul}
\usepackage{xcolor}
\usepackage{dblfloatfix}

\usepackage{pifont}
\newcommand\mynuma[1]{\ifcase#1 \or \ding{172}\or \ding{173}\or
  \ding{174}\or \ding{175}\or \ding{176}\or \ding{177}%
  \or \ding{178}\or \ding{179}\or \ding{180}\or \ding{181}\else *\fi\relax}
\newcommand\mynumb[1]{\ifcase#1 \or \ding{182}\or \ding{183}\or
  \ding{184}\or \ding{185}\or \ding{186}\or \ding{187}%
  \or \ding{188}\or \ding{189}\or \ding{190}\or \ding{191}\else *\fi\relax}

\copyrightyear{2022}
\acmYear{2022}
\setcopyright{acmcopyright}\acmConference[ICCAD '22]{IEEE/ACM International
Conference on Computer-Aided Design}{October 30-November 3, 2022}{San Diego,
CA, USA}
\acmBooktitle{IEEE/ACM International Conference on Computer-Aided Design (ICCAD
'22), October 30-November 3, 2022, San Diego, CA, USA}
\acmPrice{15.00}
\acmDOI{10.1145/3508352.3549380}
\acmISBN{978-1-4503-9217-4/22/10}

\begin{document}

\title{RT-NeRF: \underline{R}eal-\underline{T}ime On-Device \underline{Ne}ural \underline{R}adiance \underline{F}ields Towards Immersive AR/VR Rendering}


\author{Chaojian Li$^*$, Sixu Li$^*$, Yang Zhao$^*$, Wenbo Zhu, and Yingyan Lin}
\thanks{$^*$Equal contribution.}
\affiliation{%
  \institution{Georgia Institute of Technology}
  \city{Atlanta}
  \state{Georgia}
  \country{USA}
}

\input{sections/0-Abstract}

\maketitle

\input{sections/1-Introducation}

\input{sections/2-Background-Motivation}

\input{sections/3-Algorithms}
\input{sections/4-Architecture}
\input{sections/5-Experiments}
\input{sections/6-Related_Works}
\input{sections/7-Conclusion}

\begin{acks}

The work is supported by the NSF CCRI program (Award number: 2016727) and the NSF RTML program (Award number: 1937592).

\end{acks}

\newpage
\bibliographystyle{ACM-Reference-Format}
\bibliography{sample-base}

\end{document}

%% file: sections/0-Abstract.tex
\begin{abstract}
Neural Radiance Field (NeRF) based rendering has attracted growing attention thanks to its state-of-the-art (SOTA) rendering quality and wide applications in Augmented and Virtual Reality (AR/VR). However, immersive real-time ($>$ 30 FPS) NeRF based rendering enabled interactions are still limited due to the low achievable throughput on AR/VR devices. To this end, we first profile SOTA efficient NeRF algorithms on commercial devices and identify two primary causes of the aforementioned inefficiency: (1) the uniform point sampling and (2) the dense accesses and computations of the required embeddings in NeRF. Furthermore, we propose RT-NeRF, which to the best of our knowledge is the first algorithm-hardware co-design acceleration of NeRF. Specifically, \textbf{on the algorithm level}, RT-NeRF integrates an efficient rendering pipeline for largely alleviating the inefficiency due to the commonly adopted uniform point sampling method in NeRF by directly computing the geometry of pre-existing points.
Additionally, RT-NeRF leverages a coarse-grained view-dependent computing ordering scheme
for eliminating the (unnecessary) processing of invisible points.  
\textbf{On the hardware level}, our proposed RT-NeRF accelerator (1) adopts a hybrid encoding scheme to adaptively switch between a bitmap- or coordinate-based sparsity encoding format for NeRF's sparse embeddings, aiming to maximize the storage savings and thus reduce the required DRAM accesses while supporting efficient NeRF decoding; and (2) integrates both a dual-purpose bi-direction adder \& search tree and a high-density sparse search unit to coordinate the two aforementioned encoding formats.
Extensive experiments on eight datasets consistently validate the effectiveness of RT-NeRF, achieving a large throughput improvement (e.g., 9.7$\times$$\sim$3,201$\times$) while maintaining the rendering quality as compared with SOTA efficient NeRF solutions.

\end{abstract}

%% file: sections/1-Introducation.tex
\section{Introduction}
\begin{figure}[t]
  \centering
  \includegraphics[width=0.9\linewidth]{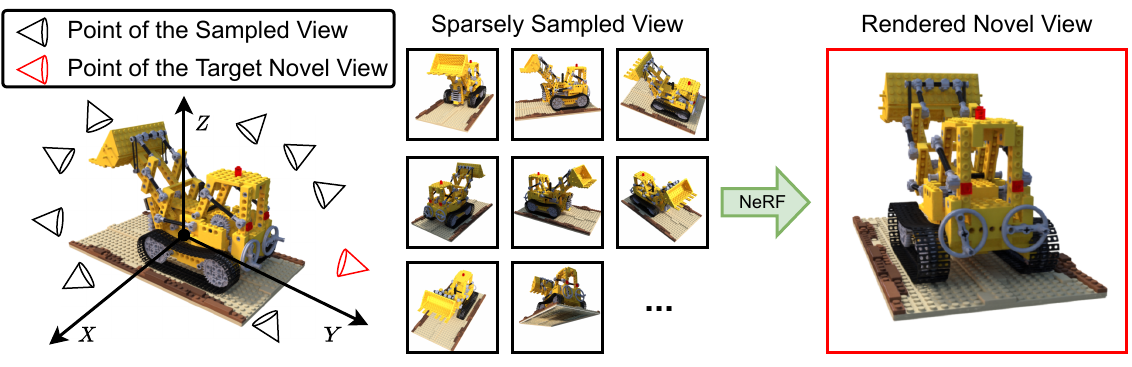}
\vspace{-1.3em}
\caption{An illustration of novel view synthesis, which is the rendering task that NeRF~\cite{mildenhall2020nerf} targets to resolve.}
  \vspace{-1.5em}
\label{fig:nvs}
\end{figure}

Novel view synthesis (see Fig.~\ref{fig:nvs}), which renders photorealistic novel views given a set of sparsely sampled views, has become a fundamental task in various AR/VR applications~\cite{dawood200919, bian2016framework,zhao2020deja,fassi2016vr,farshid2018go}, such as virtual meetings~\cite{workrooms}. 
As such, significant efforts have been made
to push forward the achievable rendering quality, among which NeRF~\cite{mildenhall2020nerf} based rendering has recently attracted much-growing attention thanks to its state-of-the-art (SOTA) rendering quality.
However, while immersive real-time ($>$ 30 FPS) NeRF based rendering enabled interactions are highly desired, they are not yet possible due to the low rendering throughput that is currently achievable on AR/VR devices, e.g., $<$ 0.04 FPS for rendering 800$\times$800 images on a SOTA GPU such as NVIDIA V100 GPUs~\cite{mildenhall2020nerf}.

To close the aforementioned gap, we first perform in-depth profiling of SOTA efficient NeRF algorithms on commercial devices by characterizing the runtime of each step in the algorithm pipeline
to identify the bottlenecks causing the rendering inefficiency. In particular, we locate two efficiency-bottleneck steps within the pipeline of SOTA efficient NeRF algorithms: (1) locating pre-existing points, which is to filter out the points representing an empty space and (2) computing pre-existing points' features, which is to generate the features (i.e., densities and colors) of pre-existing points based on the embeddings corresponding to a specific grid. Furthermore, we identify that these two steps' bottleneck inefficiencies are respectively due to (1) the commonly adopted uniform point sampling method despite the sparsities of the existed points and (2) the required dense accesses and computations of the embeddings corresponding to a specific grid, despite the sparsities of those embeddings. 

To tackle the identified inefficiencies above, we advocate NeRF algorithm-hardware co-design to achieve real-time on-device NeRF processing towards immersive AR/VR rendering, and make the following contributions:

\begin{itemize}
    \item We comprehensively profile and analyze the throughput bottlenecks in SOTA efficient NeRF-based methods on multiple commercial devices.
    We identify that (1) the commonly used uniform point sampling method and (2) the required dense accesses and computations for the embeddings are the primary causes of existing methods' inefficiencies. 
    \item We propose RT-NeRF, which to the best of our knowledge is the first algorithm-hardware co-design framework for accelerating NeRF. Specifically, our RT-NeRF aims to resolve the aforementioned bottleneck inefficiencies by developing tailored algorithm and hardware innovations that leverage the sparsities of both pre-existing points and specific grids' embeddings. Thus, our RT-NeRF has opened up an exciting perspective towards real-time NeRF solutions. 
    
    \item On the algorithm level, by leveraging the sparsities of pre-existing points, RT-NeRF integrates an efficient rendering pipeline to alleviate the inefficiency due to commonly adopted uniform point sampling by directly computing the geometry of pre-existing points based on the corresponding non-zero cubes of the occupancy grid. Additionally, in our proposed rendering pipeline, to skip the invisible points from the pre-existing ones for further boosted efficiency, RT-NeRF leverages a coarse-grained view-dependent rendering ordering scheme to avoid processing invisible points. 
    
   \item On the hardware level, our RT-NeRF accelerator adopts a hybrid encoding scheme to adaptively switch between a bitmap- or coordinate-based sparsity encoding format for NeRF embeddings with low ($<$80\%) and high ($\geq$80\%) sparsity-ratios, respectively. Such a hybrid scheme is to maximize the storage savings and thus reduce the required DRAM accesses while supporting efficient decoding, despite the diverse sparsity ratio of NeRF embeddings (e.g., 4\% $\sim$ 92\%).
   Furthermore, to avoid the commonly observed computation idleness due to sparse decoding, our RT-NeRF accelerator integrates both a high-density sparse query unit and a dual-purpose bi-direction adder\&query tree to coordinate the two aforementioned encoding formats for ensuring efficient sparse decoding.
    \item Benchmarking experiments and ablation studies on eight datasets of Synthetic-NeRF~\cite{mildenhall2020nerf} consistently validate the effectiveness of RT-NeRF, e.g., achieving 9.7$\times$ $\sim$ 3,201$\times$ throughput improvement while maintaining a similar rendering quality as compared to SOTA efficient NeRF solutions.
\end{itemize}

%% file: sections/2-Background-Motivation.tex
\section{Background and Motivation}
\label{sec:background_motivations}
\subsection{Preliminaries of NeRF}
\label{sec:nerf_preliminaries}
\begin{figure}[!t]
  \centering
  \includegraphics[width=0.9\linewidth]{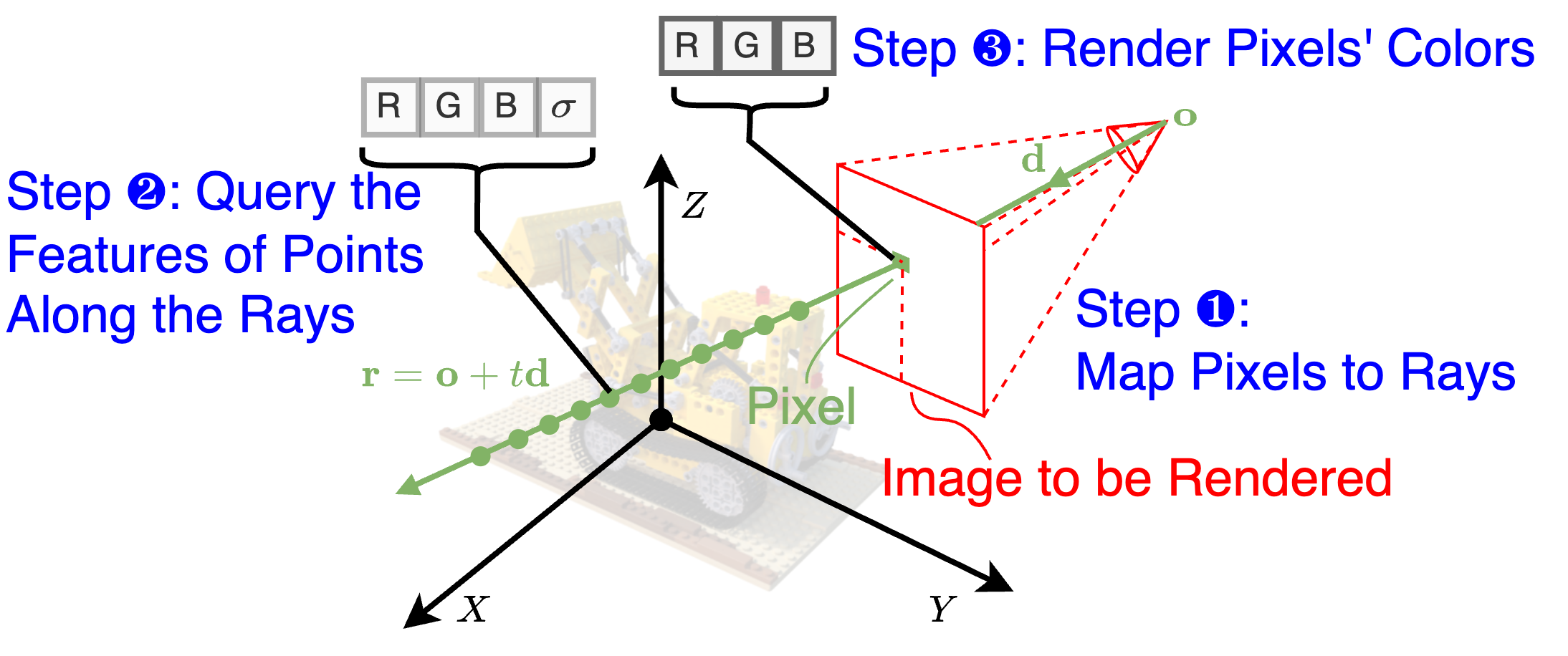}
\vspace{-1.5em}
\caption{NeRF~\cite{mildenhall2020nerf} based rendering includes Step
 {\color{black}{\ding{182}}} Map pixels to rays $\mathbf{r} = \mathbf{o}+t\mathbf{d}$ by marching camera rays through the scene, Step {\color{black}{\ding{183}}} Query the features (i.e., the RGB color and the density $\sigma$) of points along the rays by inputting their locations and distance to an MLP model, and Step \color{black}{\ding{184}} Render pixels' colors.}
\label{fig:nerf_pipeline}
 \vspace{-2.2em}
\end{figure}

\textbf{NeRF for Novel View Synthesis.}
To render photorealistic novel views as shown in Fig.~\ref{fig:nvs}, NeRF~\cite{mildenhall2020nerf} encodes a continuous volumetric field of points, which block and emit light rays, within the parameters of a multilayer perceptron (MLP). Fig.~\ref{fig:nerf_pipeline} illustrates NeRF's rendering process, which involves three steps. \textbf{Step {\color{black}{\ding{182}}}} Map pixels to rays: For each pixel to be rendered in the target novel view, a ray $\mathbf{r} = \mathbf{o}+t\mathbf{d}$ is emitted from the origin (e.g., the camera's center) of the target novel view $\mathbf{o}$ along the direction $\mathbf{d}$ to pass through this particular pixel, where $t$ represents the distance between the sampled points along this ray and the origin is denoted as $\mathbf{o}$; \textbf{Step {\color{black}{\ding{183}}}} Query the features of points along the rays: For each point that has a distance $t_k$ from $\mathbf{o}$, both its location $\mathbf{o}+t_k\mathbf{d}$ and direction $\mathbf{d}$ are sent as inputs to the MLP model $(\mathbf{o}+t_k\mathbf{d}, \mathbf{d}) \rightarrow (\sigma_k, \mathbf{c}_k)$, which outputs the corresponding density $\sigma_k$ and an RGB color $\mathbf{c}_k$ as the extracted feature of this particular point; and \textbf{Step {\color{black}{\ding{184}}}} Render pixels' colors: Following the principles of classical volume rendering~\cite{max1995optical}, the color $\mathbf{C}(\mathbf{r})$ of the pixel corresponding to the ray $\mathbf{r}$ can be computed by integrating the features of the points along the ray, which can be represented as: 

\vspace{-1.3em}
\begin{align}
\mathbf{C}(\mathbf{r}) = \sum_{k=1}^NT_k(1-\exp(-\sigma_k (t_{k+1}-t_{k})))\mathbf{c}_k, \nonumber\\
\vspace{-3em}
\textrm{ where } T_k = \exp(-\sum_{j=1}^{k} \sigma_{j} (t_{j+1}-t_{j})),
\label{eq:nerf_render}
\end{align}
where $N$ represents the number of sampled points along the ray $\mathbf{r}$ and $T_k$ denotes the accumulated transmittance along the ray $\mathbf{r}$ to the point $\mathbf{o}+t_k\mathbf{d}$, which represents the probability of the ray traveling to the point without hitting any other points. To render an image with an resolution of $H \times W$, the above steps {\color{black}{\ding{182}}} $\sim$ {\color{black}{\ding{184}}} will be repeated for $H \times W$ times, corresponding to $H \times W  \times N$ number of queries to the MLP model. As such, if using (1) an MLP of 1 million FLOPs for each query to render an image of $800 \times 800$ resolution and (2) 192 sampled points along each ray~\cite{mildenhall2020nerf} during the rendering process, the required total FLOPs would become as large as 800$\times$800$\times$192 $\times$1 million FLOPs $=$ 117 trillion FLOPs, resulting in $<$ 0.04 FPS on an NVIDIA V100 GPU~\cite{mildenhall2020nerf}. 

To alleviate the prohibitive rendering FLOPs mentioned above, various techniques have been proposed to accelerate NeRF. One popular type of approaches~\cite{chen2022tensorf,yu2021plenoctrees,sun2021direct} is to add a 3D grid, which represents the embeddings of the specific pre-set points, to be optimized together with the MLP model or even replacing the MLP model. Among them, TensoRF~\cite{chen2022tensorf} has achieved the SOTA efficiency in terms of accuracy vs. the-number-of-parameters trade-offs, which makes it possible to be further compressed for being deployed on resource-constrained AR/VR devices, e.g., the popular Oculus Quest 2 VR headset has only $<$ 6GB RAM and $<$ 14 watt-hour battery capacity~\cite{quest2}. It is worth noting that another type of approaches accelerates NeRF by caching a large amount of intermediate results, e.g., $>$ 54 GB in FastNeRF~\cite{garbin2021fastnerf}, and thus can only be used in high-end GPUs with sufficient resources instead of resource-constrained AR/VR devices that our proposed RT-NeRF mainly targets.

\begin{figure}[!b]
\vspace{-1.em}
  \centering
  \includegraphics[width=1.0\linewidth]{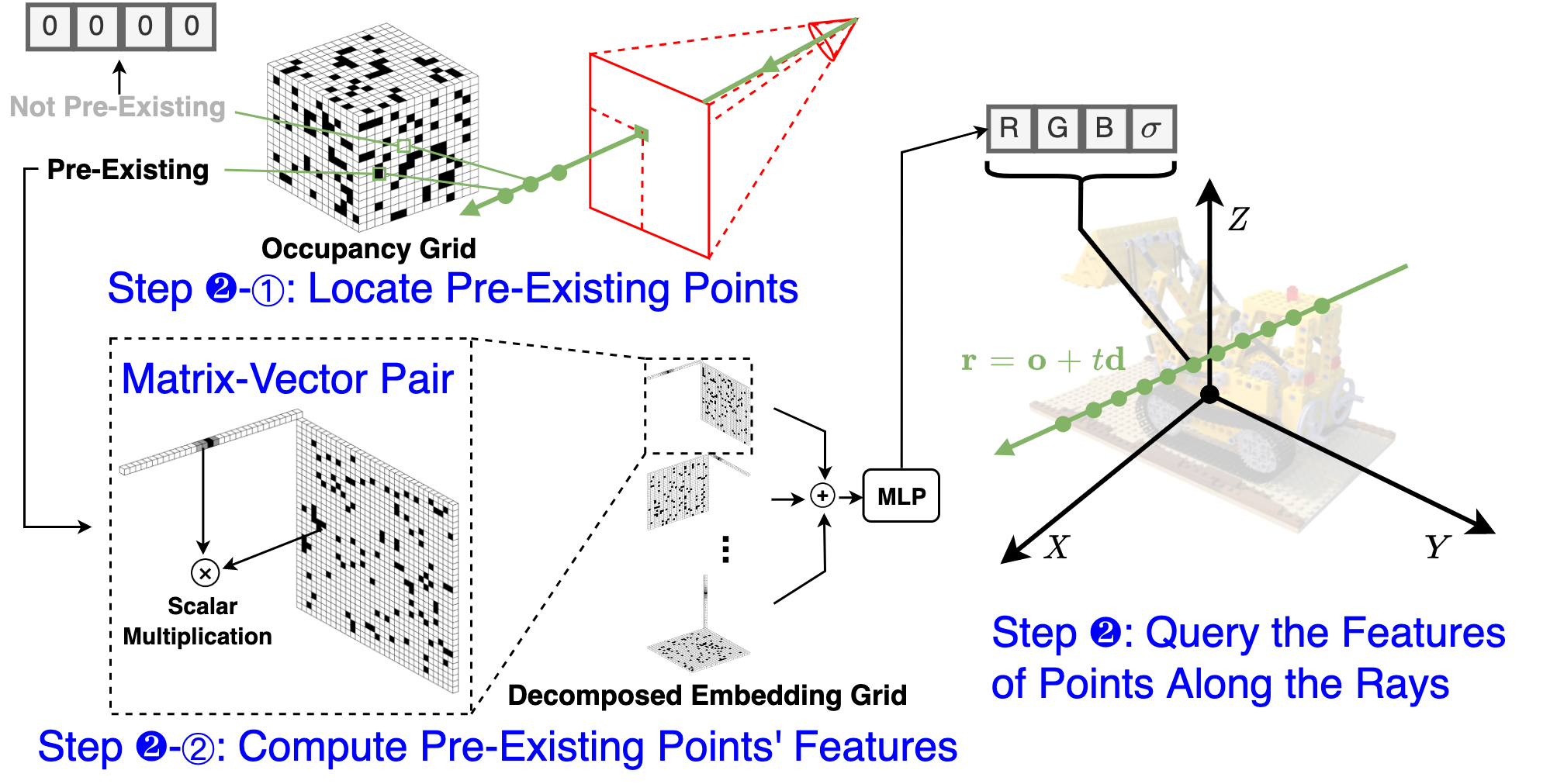}
\vspace{-2.5em}
\caption{TensoRF~\cite{chen2022tensorf} achieving SOTA NeRF efficiency replaces Step {\color{black}{\ding{183}}} (i.e., query the features of points along the rays using a MLP) in NeRF~\cite{mildenhall2020nerf} with both Step {\color{black}{\ding{183}}}-{\color{black}{\ding{172}}}, which locates pre-existing points using an occupancy grid, and Step {\color{black}{\ding{183}}}-{\color{black}{\ding{173}}}, which computes pre-existing points' features based on a decomposed embedding grid in terms of matrix-vector pairs.}
\label{fig:tensorf_pipeline}
 \vspace{-0.5em}
\end{figure}

\begin{figure*}[t]
  \centering
  \includegraphics[width=1.0\linewidth]{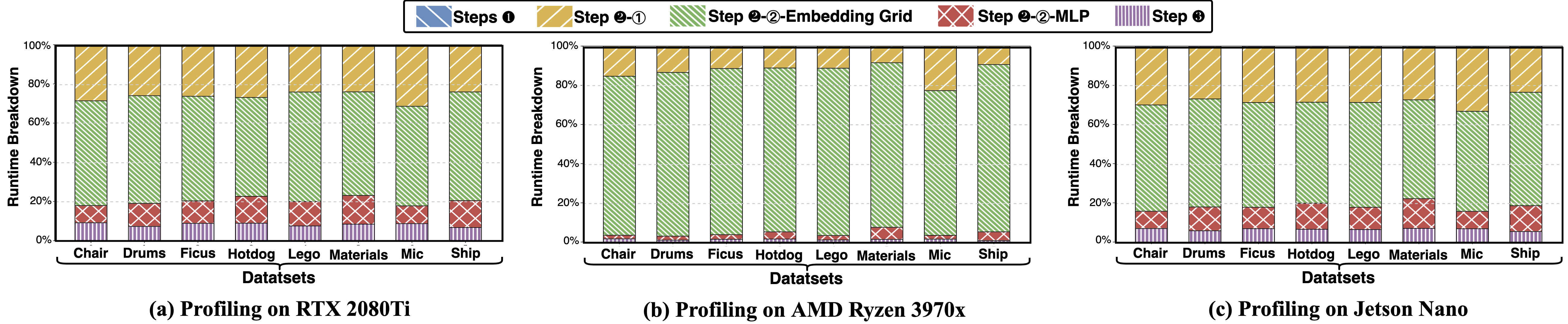}
\vspace{-2.5em}
\caption{Runtime breakdown across eight datasets on three representative commercial devices, which shows that among Step {\color{black}{\ding{182}}} (i.e., map pixels to rays), Step {\color{black}{\ding{183}}}-{\color{black}{\ding{172}}} (i.e., locate the pre-existing points), Step {\color{black}{\ding{183}}}-{\color{black}{\ding{173}}} (i.e., compute pre-existing points' features), and Step {\color{black}{\ding{184}}} (i.e., render pixels' colors), the SOTA efficient NeRF solution~\cite{chen2022tensorf} is bottlenecked by Step {\color{black}{\ding{183}}}-{\color{black}{\ding{172}}} and Step {\color{black}{\ding{183}}}-{\color{black}{\ding{173}}}, the latter of which includes Step {\color{black}{\ding{183}}}-{\color{black}{\ding{173}}}-Embedding-Grid and Step {\color{black}{\ding{183}}}-{\color{black}{\ding{173}}}-MLP that correspond to the operations in Eq.~\ref{eq:tensorf_decomp} and for the MLP inference, respectively).}
  \vspace{-1.3em}
\label{fig:profile_tensorf}
\end{figure*}

\textbf{TensoRF with SOTA Efficiency.}
As visualized in Fig.~\ref{fig:tensorf_pipeline}, TensoRF~\cite{chen2022tensorf}, which currently represents SOTA NeRF efficiency, replaces Step {\color{black}{\ding{183}}} (i.e., query the features of points along the rays using a MLP) in NeRF~\cite{mildenhall2020nerf} with the following two steps. \textbf{Step {\color{black}{\ding{183}}}-{\color{black}{\ding{172}}}} Locate pre-existing points: For each point with a distance $t_k$ to the target novel view $\mathbf{o}$, its indices $(x_k, y_k, z_k) \in \mathbb{N}^3$ of a 3D binary occupancy grid can be computed by quantizing its coordinates $\mathbf{o}+t_k\mathbf{d} \in \mathbb{R}^3$. If the corresponding value in the occupancy grid is zero, i.e., this point is in an empty space and thus does not exist, TensoRF returns zero as the corresponding features and skips computing the following steps. \textbf{Step {\color{black}{\ding{183}}}-{\color{black}{\ding{173}}}} Compute pre-existing points' features: For each pre-existing point identified in the previous step, its embeddings are queried from a 3D grid of the specific pre-set points' embeddings, i.e., the embedding grid, based on the aforementioned indices $(x_k, y_k, z_k)$. To save the cost of storing and accessing such an embedding grid, TensoRF~\cite{chen2022tensorf} further decomposes the embedding grid into multiple matrix-vector pairs. Thus, the corresponding density $\mathbf{\sigma}_k$ can be computed by: 

\vspace{-1.5em}
\begin{align}
\mathbf{\sigma}_k = \sum_{r=1}^R(&\mathbf{v}_{r, x_k}^X\cdot \mathbf{M}_{r, (y_k, z_k)}^{Y,Z}+\mathbf{v}_{r, y_k}^Y\cdot \mathbf{M}_{r, (x_k, z_k)}^{X,Z}+\mathbf{v}_{r, z_k}^Z\cdot \mathbf{M}_{r, (x_k, y_k)}^{X,Y}),
\label{eq:tensorf_decomp}
\end{align}
\vspace{-1.3em}

\hspace{-1.5em} where $R$ represents the number of matrix-vector pairs in three decomposition modes, i.e., the 3D embedding grid is decomposed into the outer product of the (1) matrices in the $Y,Z$ plane (i.e., $\mathbf{M}^{Y,Z}$) and vectors along the $X$ axis (i.e., $\mathbf{v}^X$), (2) matrices in the $X,Z$ plane (i.e., $\mathbf{M}^{X,Z}$) and vectors along the $Y$ axis (i.e., $\mathbf{v}^Y$), and (3) matrices in the $X,Y$ plane (i.e., $\mathbf{M}^{X,Y}$) and vectors along the $Z$ axis (i.e., $\mathbf{v}^Z$). Thus, $\mathbf{v}_{r, x_k}^X$ denotes the $x_k$-th element of the vector along the $X$ axis in the $r$-th matrix-vector pair, the subscripts of $\mathbf{v}_{r, y_k}^Y$, $\mathbf{v}_{r, z_k}^Z$, $\mathbf{M}_{r, (y_k, z_k)}^{Y,Z}$, $\mathbf{M}_{r, (x_k, z_k)}^{X,Z}$, $\mathbf{M}_{r, (x_k, y_k)}^{X,Y}$ can be interpreted in the same way. Meanwhile, the corresponding color $\mathbf{c}_k$ (see Eq.~\ref{eq:nerf_render}) can be accessed in the same aforementioned way from another set of matrix-vector pairs. Additionally, 
a small MLP takes both (1) the concatenated results of the scalar multiplication among different matrix-vector pairs and (2) the direction $\mathbf{d}$ as its inputs to generate the view-dependent color for the next steps.
Note that TensoRF~\cite{chen2022tensorf} also adopts  early-ray-termination~\cite{kruger2003acceleration} to filter out invisible points from  pre-existing points when computing colors. Specifically, pre-existing points with an accumulated transmittance $T_k$ that is smaller than a pre-set threshold nearly do not contribute to the final rendered color $\mathbf{C}(\mathbf{r})$ as suggested in Eq.~\ref{eq:nerf_render}, and thus can be regarded as invisible and the corresponding computations for generating their colors can be skipped.

\subsection{Profile SOTA Efficient NeRF Solutions}
To better understand the throughput bottleneck of SOTA efficient NeRF solutions, we analyze the runtime breakdown of each step within TensoRF~\cite{chen2022tensorf}'s rendering pipeline on three representative commercial devices, including RTX 2080Ti~\cite{2080ti} (a GPU for cloud computing), AMD Threadripper 3970x~\cite{ryzen} (a CPU for cloud computing), and Jetson Nano~\cite{jetson_nano} (an embedded GPU for edge computing). As shown in Fig.~\ref{fig:profile_tensorf}, the profiling results on the eight commonly used datasets of Synthetic-NeRF~\cite{mildenhall2020nerf} show that Step {\color{black}{\ding{183}}}-{\color{black}{\ding{172}}} (i.e., locate the pre-existing points) and Step {\color{black}{\ding{183}}}-{\color{black}{\ding{173}}} (i.e., compute those pre-existing points' features) dominate the overall rendering latency of TensoRF~\cite{chen2022tensorf} on all these commercial devices that target both cloud and edge computing.

\subsubsection{Analyze Step {\color{black}{\ding{183}}}-{\color{black}{\ding{172}}}} \hfill \break
\label{sec:background_locating_existing_points}
\textbf{Identified Causes of Inefficiency.} As shown in Fig.~\ref{fig:tensorf_pipeline}, to locate pre-existing points, all the candidate points are uniformly sampled along rays and then the existence of pre-existing points are identified via a query process based on the occupancy grid. From this process, we identify two sources of redundant costs: (1) the sparsity of the occupancy grid is not leveraged, and thus the number of queries to the occupancy grid is fixed as $H \times W \times N$ regardless of the values in the occupancy grid, where $H$ and $W$ represent the height and width of the image to be rendered, respectively, and $N$ denotes the number of sampled points along each ray; and (2) the DRAM accesses to the occupancy grid are irregular because the emitted rays can come from any direction, and thus the order of their accesses to the occupancy grid can not be predicted in advance.

\textbf{Proposed Solution.} To eliminate the above redundant computations and memory accesses, we propose an efficient NeRF pipeline, which directly computes the coordinates of pre-existing points by looping over the non-zero cubes of the occupancy grid instead of all the sampled candidate points. The detailed description of our proposed pipeline is provided in Sec.~\ref{sec:method_alg_computation_pipeline}.

\begin{figure*}[t]
  \centering
  \includegraphics[width=1.0\linewidth]{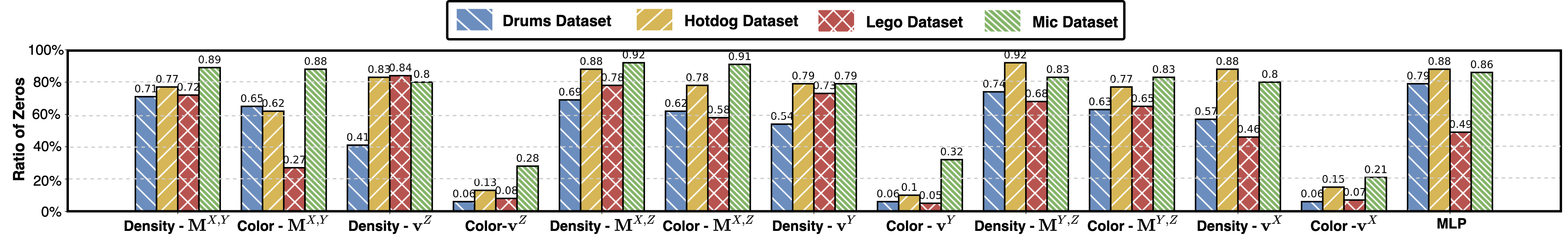}
\vspace{-2.5em}
\caption{The sparsity of different weights in Eq.~\ref{eq:tensorf_decomp} on the Drums, Hotdog, Lego, and Mic datasets, where Density - $\mathbf{M}^{X,Y}$ represents the matrices in the $X$, $Y$ plane for densities and the notations of the other weights can be interpreted in the same way.}
  \vspace{-1.em}
\label{fig:sparsity}
\end{figure*}

\subsubsection{Analyze Step {\color{black}{\ding{183}}}-{\color{black}{\ding{173}}}} \hfill \break
\label{sec:background_computing_existing_points_features}
\textbf{Identified Cause of Inefficiency.}
As illustrated in Fig.~\ref{fig:profile_tensorf}, for Step {\color{black}{\ding{183}}}-{\color{black}{\ding{173}}} (i.e., compute pre-existing points' features), the required latency of querying the embeddings from the decomposed embedding grid in the format of matrix-vector pairs is much higher (e.g., 4 $\times$ $\sim$ 45 $\times$) than that of computing the view-dependent colors using the MLP model. We identify that this is because the matrix-vector pairs are treated as dense matrices and/or vectors, causing both redundant computations and DRAM accesses despite their high sparsity (e.g., up to 92\% sparsity) as visualized in Fig.~\ref{fig:sparsity}.

\textbf{Proposed Solution.} 
Looking into the aforementioned step of computing pre-existing points' features,  
we find that there consistently exist sparsities in the corresponding weights, and these sparsities feature imbalanced patterns and are dataset-dependent, which however have not been leveraged by SOTA efficient NeRF solutions. 
As shown in Fig.~\ref{fig:sparsity}, regarding the aspect of \ul{imbalanced sparsity patterns}, we can observe that the sparsity ratio of different types of weights can range from 4\% to 92\%; regarding the aspect of \ul{dataset-dependent} sparsity, the sparsity ratio of the same type of weights can range from 46\% to 88\% across different datasets. To utilize the aforementioned sparsities for boosted efficiency, we propose a hybrid encoding scheme for the matrix and/or vectors that adaptively adopts a bitmap- or coordinate-based sparsity encoding format for low ($<$80\%) and high ($\ge$80\%) sparsity-ratio scenarios, respectively, aiming to maximize the storage savings and thus reduce the required DRAM accesses. Additionally, we propose a high-density sparse query unit and a dual-purpose bi-direction adder\&query tree to coordinate these two encoding
formats above for ensuring efficient sparse decoding. The detailed descriptions of the proposed hybrid encoding scheme, sparse query unit, and adder\&query tree are provided in Sec.~\ref{sec:HWSparse}.

%% file: sections/3-Algorithms.tex
\section{RT-NeRF: Proposed Algorithm}
\label{sec:algorithm}
\begin{figure}[!b]
  \vspace{-2em}
  \centering
  \includegraphics[width=1.0\linewidth]{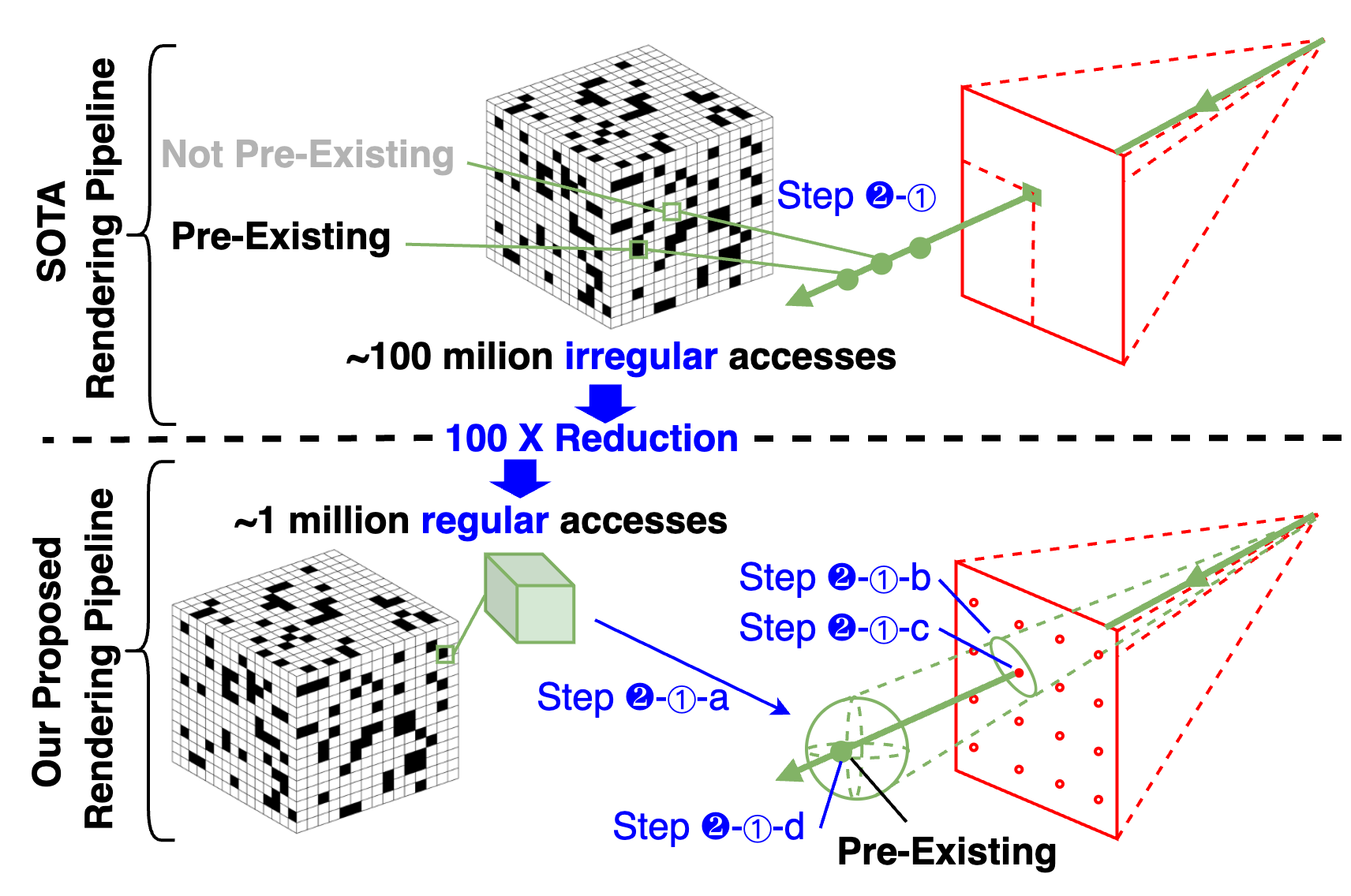}
  \vspace{-2.5em}
\caption{The proposed rendering pipeline which directly computes the geometry of pre-existing points, enabling occupancy grid accesses that are both \textit{fewer} and \textit{more regular} than the SOTA rendering pipeline.}
\label{fig:algs}
\end{figure}

\subsection{Efficient Rendering Pipeline}
\label{sec:method_alg_computation_pipeline}
To alleviate redundant computations and memory accesses due to commonly adopted uniform point sampling in Step {\color{black}{\ding{183}}}-{\color{black}{\ding{172}}} (i.e., locate pre-existing points) as analyzed in Sec.~\ref{sec:background_locating_existing_points}, 
we propose an efficient rendering pipeline, which directly computes the geometry of pre-existing points by looping over only the non-zero cubes of the occupancy grid instead of all the sampled candidate points. As demonstrated in Fig.~\ref{fig:algs}, compared to the SOTA rendering pipeline in~\cite{chen2022tensorf}, the proposed one can reduce the number of accesses to the occupancy grid by 100 $\times$ and also makes the corresponding DRAM accesses more regular, i.e., looping over the non-zero cubes in the occupancy grid with a fixed order instead of randomly accessing the grid based on the unpredictable directions of rays. Specifically, the proposed rendering pipeline consists of \textbf{Step {\color{black}{\ding{183}}}-{\color{black}{\ding{172}}}-a}: Approximate each non-zero cube in the occupancy grid as a ball for ease of computations in the following steps; \textbf{Step {\color{black}{\ding{183}}}-{\color{black}{\ding{172}}}-b}: Project the aforementioned ball to the image to be rendered as an oval; \textbf{Step {\color{black}{\ding{183}}}-{\color{black}{\ding{172}}}-c}: Identify the points inside the oval based on the regular arrangement of the points in the image to be rendered, i.e., one point corresponds to one pixel; \textbf{Step {\color{black}{\ding{183}}}-{\color{black}{\ding{172}}}-d}: Solve the geometry of the points that are both (1) inside the ball and (2) along the rays passing through the points inside the oval, using an analytic solution of line–sphere intersections~\cite{eberly20063d}. Thus, in the proposed rendering pipeline, only the pre-existing points will be included in the loop instead of all the sampled candidate points, resolving the limitations of (1) ignoring the sparsity of the occupancy grid and (2) requiring irregular DRAM accesses in the SOTA 
rendering pipeline~\cite{chen2022tensorf} as analyzed in Sec.~\ref{sec:background_locating_existing_points}.

\subsection{View-Dependent Rendering Ordering } \hfill \break
\label{sec:HWTiling}
\begin{figure}[b]
\vspace{-2.4em}
  \centering
\includegraphics[width=0.9\linewidth]{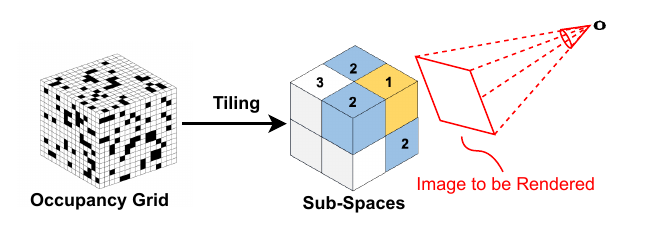}
\vspace{-1.5em}
\caption{The tiled sub-space (marked as yellow) that is closest to the origin of the target view $\mathbf{o}$ will be processed first during our rendering process based on the current target view.}
\label{fig:proalgpattern}
\end{figure}
As illustrated in Sec.~\ref{sec:nerf_preliminaries}, early-ray-termination~\cite{kruger2003acceleration}, which can filter out invisible points from pre-existing points, has been widely adopted in SOTA efficient NeRF solutions~\cite{chen2022tensorf,yu2021plenoctrees,hedman2021baking}. However, applying early-ray-termination~\cite{kruger2003acceleration} to our proposed efficient rendering pipeline in Sec.~\ref{sec:method_alg_computation_pipeline} is not straightforward. This is because a given point's visibility is dependent on the features of the points that are closer to the view origin as suggested in Eq.~\ref{eq:nerf_render}, where a lower accumulated transmittance $T_k$ indicates lower visibility. Thus, there exists redundant computations and data accesses in Step {\color{black}{\ding{183}}}-{\color{black}{\ding{173}}} (i.e., compute pre-existing points' features) 
if those invisible but pre-existing points are accessed first while the corresponding accumulated transmittance is still unknown because of the lack of features for points that are closer to the view origin than those invisible points.

To close the gap above, we propose a coarse-grained view-dependent \sloppy rendering order. Specifically, as shown in Fig.~\ref{fig:proalgpattern}, the occupancy grid is first tiled into eight sub-spaces and the non-zero cubes in the sub-space that is closest to the origin of the target view will enter Step {\color{black}{\ding{183}}}-{\color{black}{\ding{172}}} (i.e., locate the pre-existing points) earlier. 
In this way, the features of points that are closer to the view origin are calculated first which can help determine the visibility of points that are more distant from the view origin and thus can prevent the processing of invisible points.  
Additionally, such firstly located pre-existing points will also enter Step {\color{black}{\ding{183}}}-{\color{black}{\ding{173}}} (i.e., compute pre-existing points' features) earlier. Therefore, only the partial sum of the final rendered color $\mathbf{C}(\mathbf{r})$ in Eq.~\ref{eq:nerf_render} needs to be stored as the intermediate results during rendering, in contrast to the queried features of all pre-existing points in SOTA solutions. Thus, the proposed coarse-grained view-dependent rendering order not only prevents unnecessary computations and memory accesses for invisible points in the steps of locating pre-existing points and computing the features of pre-exisitng points, but also reduces the memory accesses in Step \color{black}{\ding{184}} (i.e., render pixels' colors) based on Eq.~\ref{eq:nerf_render}, effectively accelerating all the steps that account for $>$ 99 \% of the rendering latency as visualized in Fig.~\ref{fig:profile_tensorf}.

%% file: sections/4-Architecture.tex
\section{RT-NeRF: Proposed Accelerator}
\label{sec:RT-NeRF_acceleration_system}
\begin{figure}[t]
  \centering
  \includegraphics[width=0.9\linewidth]{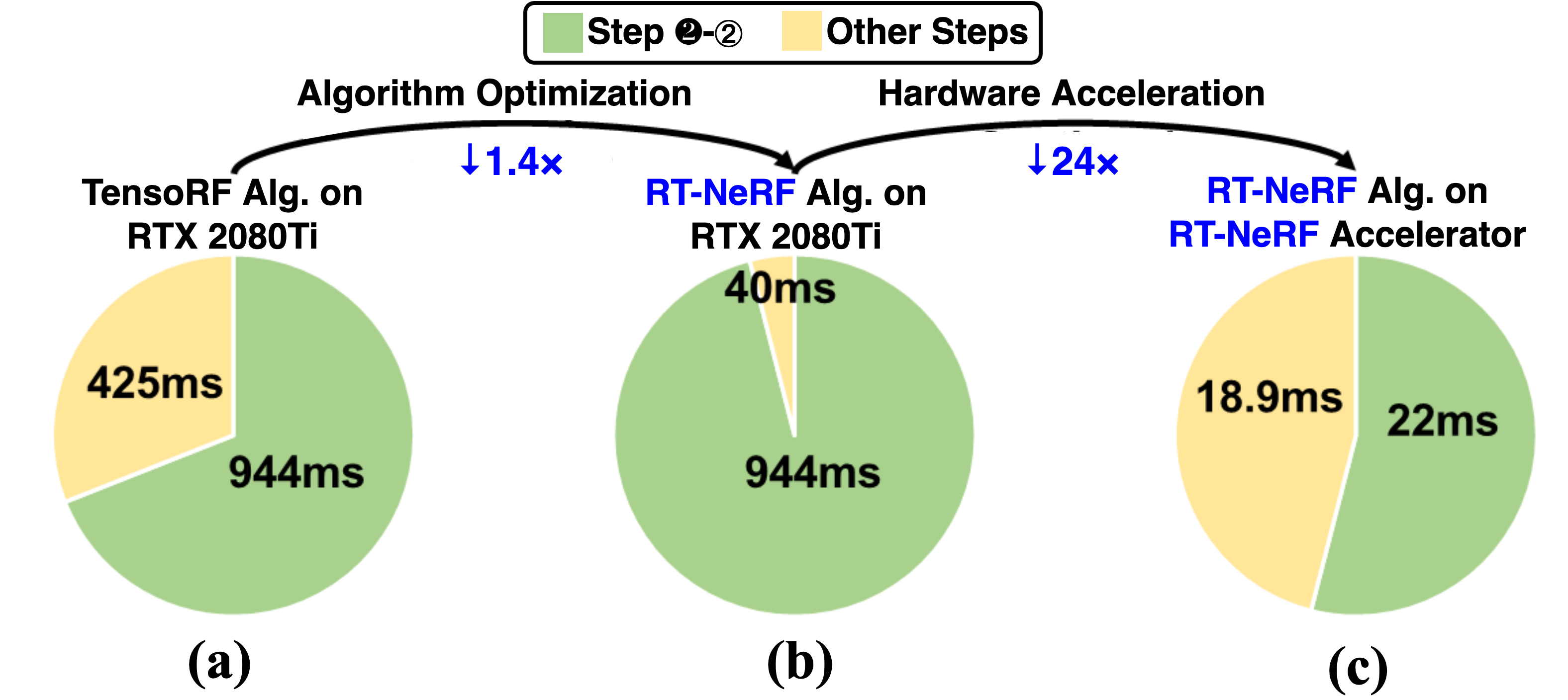}
\vspace{-1.em}
\caption{The resulting changes in the runtime breakdown (averaged over eight datasets of Synthetic-NeRF~\cite{mildenhall2020nerf}) of \textbf{Step {\color{black}{\ding{183}}}-{\color{black}{\ding{173}}}} (i.e., compute pre-existing points' features) and other steps, after applying our proposed algorithm optimization and hardware acceleration introduced in Sec.~\ref{sec:algorithm} and Sec.~\ref{sec:RT-NeRF_acceleration_system}, respectively.}
\vspace{-2.em}
\label{fig:technique}
\end{figure}

\textbf{Motivation.} As shown in Fig.~\ref{fig:technique} (a) and (b), when executing our RT-NeRF algorithm (see Sec.~\ref{sec:algorithm}) on commercial devices, Step {\color{black}{\ding{183}}}-{\color{black}{\ding{173}}} (i.e., compute pre-existing points' features) now becomes the only throughput bottleneck, which can cost 96\% of the total rendering latency. This set of profiling results (1) verify the effectiveness of both our proposed efficient rendering pipeline and view-dependent rendering ordering method described in Sec.~\ref{sec:method_alg_computation_pipeline} and Sec.~\ref{sec:HWTiling}, respectively, e.g., $\downarrow$1.4$\times$ rendering latency reduction as compared to TensoRF~\cite{chen2022tensorf}; and (2) suggest further accelerating the step of computing pre-existing points' features by leveraging the sparsity of the matrix-vector pairs as analyzed in Sec.~\ref{sec:background_computing_existing_points_features}. To this end, we propose a dedicated RT-NeRF accelerator to take advantage of the sparsity in the matrix-vector pairs to further accelerate the rendering process, e.g., reducing the rendering latency by $\downarrow$24$\times$ as compared to that of the RTX 2080Ti GPU when running the same algorithm, as visualized in Fig.~\ref{fig:technique} (c).

In this section, we first analyze the design challenges in leveraging the sparsity of the aforementioned matrix-vector pairs in Sec.~\ref{sec:after_challenge}, and then present our proposed RT-NeRF accelerator in Sec.~\ref{sec:hardware}.

\subsection{Unique Challenges}
\label{sec:after_challenge}
Benefiting from the inherent sparsity in the matrix-vector pairs of decomposed embedding grids (see Fig.\ref{fig:tensorf_pipeline}), the corresponding computations and memory accesses can be skipped to minimize the rendering latency. However, the sparsity ratios of different matrices and vectors feature \textit{imbalance} patterns and are \textit{dataset-dependent}, e.g., ranging from 4\% to 92\% among different matrices and vectors even within one dataset as analyzed in Sec.~\ref{sec:background_computing_existing_points_features}. Such imbalanced and dataset-dependent sparsity makes it difficulty, i.e., \textbf{the first challenge}, to efficiently \textbf{encode} the sparse matrices and vectors for minimizing both the required storage size and thus DRAM accesses. Furthermore, since decoding the sparse metadata (i.e., information about the indices of non-zero elements) can require several processing cycles on the metadata and thus additional latency, computation idleness may be introduced due to the necessity of waiting for this decoding processing~\cite{dave2021hardware}. Therefore, \textbf{the second challenge} for the accelerator design is to ensure efficient \textbf{decoding} of the sparse metadata to prevent potentially introduced computation idleness. 
   
To tackle the above two challenges, we propose a dedicated accelerator which (1) adopts a hybrid encoding scheme to adaptively switch between a bitmap- and coordinate-based encoding format for NeRF's matrices and vectors of low sparsity-ratio (<80\%) and high (>80\%) sparsity-ratio, respectively, which is to minimize their memory storage size and thus the required DRAM accesses~\cite{dave2021hardware} while supporting efficient decoding; and (2) integrates two efficient decoding\&search modules including both a high-density sparse search unit and a dual-purpose bi-direction adder \& search tree to implement efficient sparse decoding for both of the two sparse encoding formats above and thus prevent computation idleness commonly observed due to sparse decoding.

\label{sec:HWOverallArch} 
\subsection{The Proposed Accelerator}
\label{sec:hardware} 
In this section, we first provide an overview of the accelerator in Sec.~\ref{sec:HWoverview}, and then present both the high-density sparse search unit and dual-purpose bi-direction adder \& search tree unit, which is to coordinate with the adopted hybrid encoding scheme (see Sec.~\ref{sec:HWSparse}) to take advantage of the sparsity in NeRF's matrices and vectors mentioned above (see Fig.~\ref{fig:sparsity}) and address the unique challenges discussed in Sec.~\ref{sec:after_challenge}. 

\begin{figure}[t]
  \centering
\includegraphics[width=1.0\linewidth]{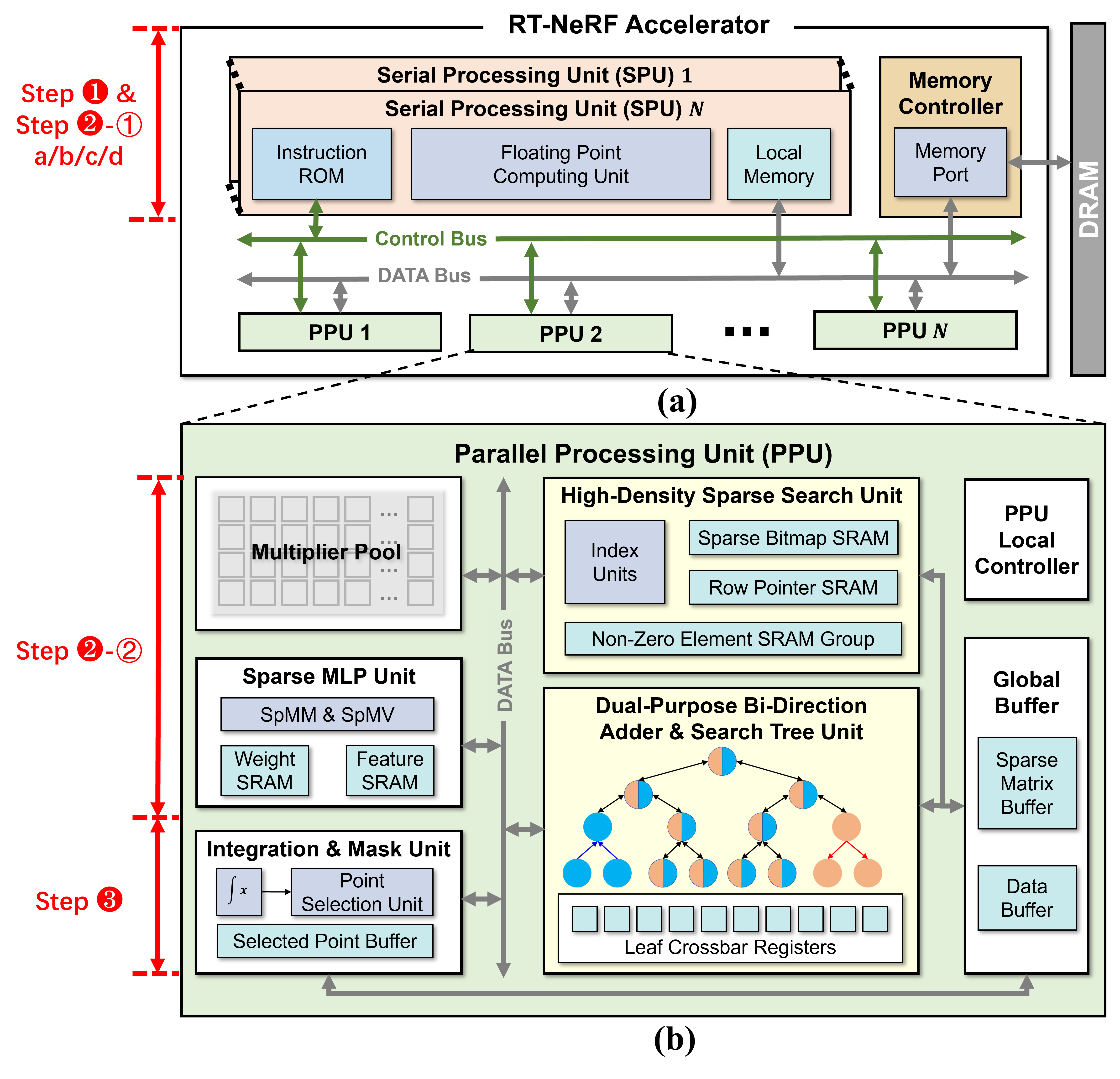}
\vspace{-2.5em}
\caption{Overall architecture of the proposed accelerator, illustrating the block diagram of both the (a) overall micro-architecture and (b) Parallel Processing Unit.}
  \vspace{-1.8em}
\label{fig:HWOverall}
\end{figure}

\subsubsection{Architecture Overview}
\label{sec:HWoverview} \hfill \break
Fig.~\ref{fig:HWOverall} (a) shows the overall architecture of our RT-NeRF accelerator, consisting of a memory controller for handling data communication with the off-chip DRAM (Fig.~\ref{fig:HWOverall} (a) top-right), $N$ Serial Processing Units (SPUs) (Fig.~\ref{fig:HWOverall} (a) top-left), and $N$ Parallel Processing Units (PPUs) (Fig.~\ref{fig:HWOverall} (a) bottom), which are demonstrated in Fig.~\ref{fig:HWOverall} (b).
Specifically, the SPUs are dedicated to Step {\color{black}{\ding{182}}} and Step {\color{black}{\ding{183}}}-{\color{black}{\ding{172}}}-a/b/c/d; while Step {\color{black}{\ding{183}}}-{\color{black}{\ding{173}}} and Step {\color{black}{\ding{184}}} is processed on the PPUs to leverage the sparsity of the matrix-vector pairs analyzed in Sec.~\ref{sec:background_computing_existing_points_features}.
\textbf{The SPU} adopts a RISC-V based processor design, the core of which is a shared floating point computing unit (FPU) for efficiently processing floating point computations, e.g., the analytic solution of line–sphere intersections~\cite{eberly20063d} for solving the geometry of pre-existing points in Step {\color{black}{\ding{183}}}-{\color{black}{\ding{172}}}-d described in Sec.~\ref{sec:method_alg_computation_pipeline}. The reason for adopting a shared FPU other than assigning separate FPUs for each step is that a shared FPU is already sufficient thanks to our proposed RT-NeRF algorithm, in which the operations to be deployed on the FPU account for only < 5\% of the total rendering computations.

\textbf{The PPUs} (see Fig.~\ref{fig:HWOverall} (b)) are to accelerate the remaining $\geq$ 95\% of computations corresponding to both Step {\color{black}{\ding{183}}}-{\color{black}{\ding{173}}} and Step {\color{black}{\ding{184}}}.
Specifically, a PPU 
consists of (1) a multiplier pool (top-left in Fig.~\ref{fig:HWOverall} (b)) for the multiplications in computing pre-existing points' features (i.e., densities and colors), (2) a sparse MLP unit (middle-left in Fig.~\ref{fig:HWOverall} (b)) for processing the MLP inference, (3) an integration \& mask unit (bottom-left in Fig.~\ref{fig:HWOverall} (b)) for computing the final rendered color by integrating the features of the points along the same ray as suggested in Eq.~\ref{eq:nerf_render}, (4) a dual-purpose bi-direction adder \& search tree unit  (bottom-middle in Fig.~\ref{fig:HWOverall} (b)) for handling the accumulations of computing pre-existing points' features and efficient metadata decoding of matrices/vectors with a high sparsity ($\geq$ 80\%), (5) a high-density sparse search unit (top-middle in Fig.~\ref{fig:HWOverall} (b)) for efficient metadata decoding of matrices/vectors with a low sparsity ($<$ 80\%), (6) memories for storing the encoding metadata, elements of the matrix-vector pairs, and the intermediate results, and (7) a local controller.

\subsubsection{Hybrid Sparse Encoding Scheme and Efficient Sparse Decoding\&Search Modules} \hfill \break
\label{sec:HWSparse} 
In this subsection, we first (1) introduce the bitmap-based encoding format for the matrices/vectors with a low sparsity ($<$ 80\%), which account for 68\% of the overall matrices/vectors across eight datasets of Synthetic-NeRF~\cite{mildenhall2020nerf}, and the high-density sparse search unit for the corresponding efficient decoding; 
and then (2) demonstrate the coordinate-based encoding format for the matrices/vectors with a high sparsity ($\geq$ 80\%), which account for 32\% of the overall matrices/vectors across eight datasets of Synthetic-NeRF~\cite{mildenhall2020nerf}, and the dual-purpose bi-direction adder \& search tree unit for efficiently handling the corresponding decoding.

\begin{figure}[t]
  \centering
  \includegraphics[width=0.85\linewidth]{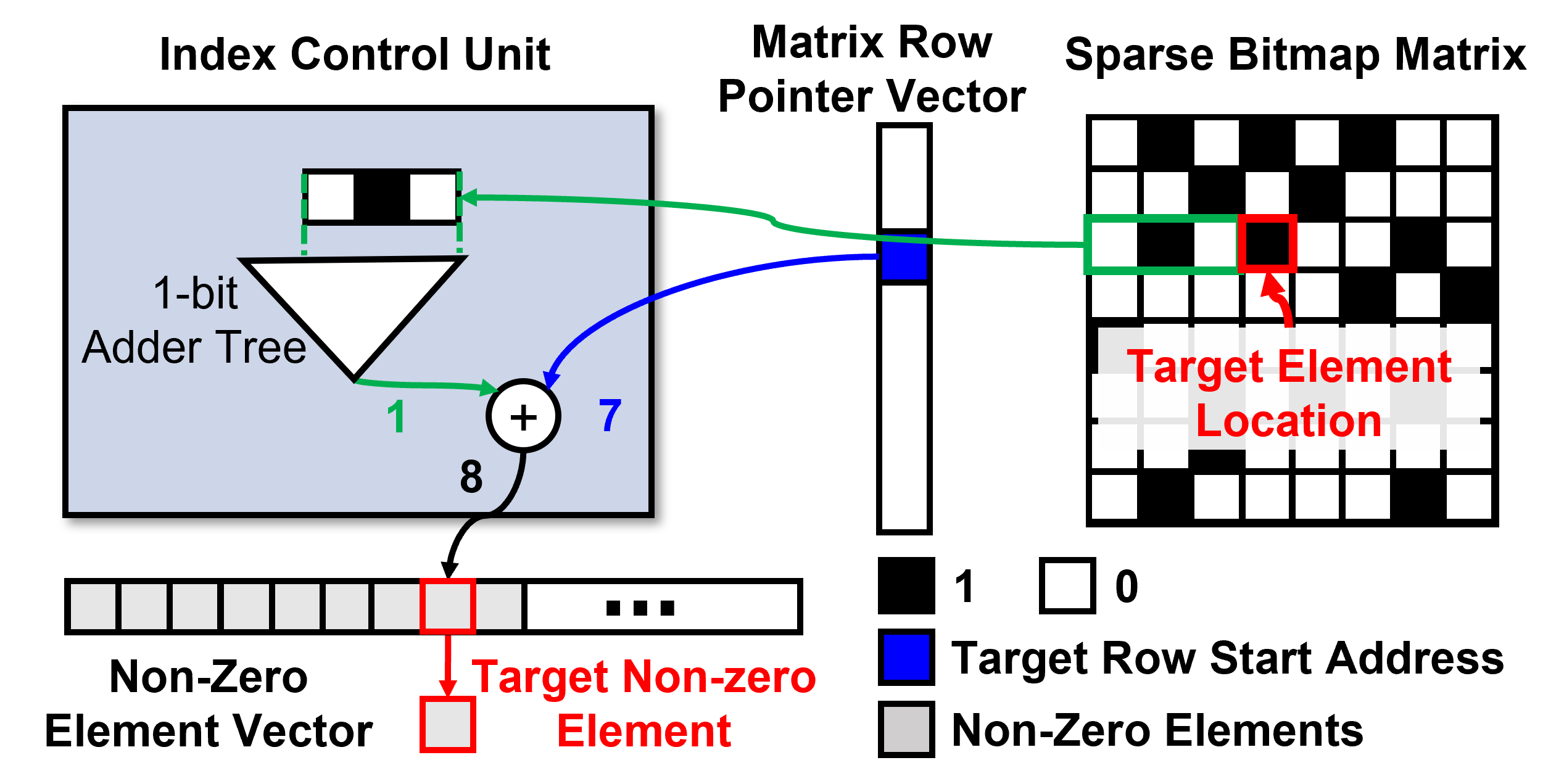}
\vspace{-1.em}
\caption{The proposed bitmap-based sparse encoding format and the high-density sparse search unit.}
\vspace{-1.8em}
\label{fig:HDSparseIndex}
\end{figure}

\textbf{Bitmap-Based Sparse Encoding Format and High-Density Sparse Search Unit.}
The bitmap encoding~\cite{dave2021hardware}, which represents the sparsity of each element in the sparse matrices/vectors as 1-bit binary metadata (i.e., 0 for zero elements and 1 for non-zero elements), is a commonly used sparse encoding method for matrices/vectors with a low sparsity.
However, directly using bitmap encoding for sparse matrices/vectors can result in varying decoding latencies, which depends on the location of the elements to be decoded, and thus introduce potential computation idleness when a decoding process with a large latency occurs.
To tackle this issue of varying decoding latencies, we propose a bitmap-based sparsity encoding format for the matrices/vectors of decomposed embedding grids. As shown in Fig.~\ref{fig:HDSparseIndex}, the proposed bitmap-based sparsity encoding format contains a newly-introduced matrix row pointer vector (middle in Fig.~\ref{fig:HDSparseIndex}), a sparse bitmap matrix (right in Fig.~\ref{fig:HDSparseIndex}), and an vector of non-zero elements (bottom-left in Fig.~\ref{fig:HDSparseIndex}). The newly-introduced matrix row pointer vector stores the addresses of the first non-zero element of each row in the sparse matrix or the start addresses of each row. 
Thanks to the proposed bitmap-based encoding format, the maximum search latency for any element location in the sparse matrix is fixed, e.g., three cycles in our specific design, eliminating the potential computation idleness for the pipelined decoding processes of multiple elements.
 
The decoding process of the proposed high-density sparse search unit is as follows: Given a target element's location $(x,y)$, \underline{Cycle 1}: the row $x$ of the sparse bitmap matrix is fetched and the 1-bit encoding metadata of $(x,y)$ is checked. If the 1-bit encoding metadata is zero, the search result is zero; otherwise, the high-density search unit executes the following steps of Cycle 2 and Cycle 3; \underline{Cycle 2}: the 1-bit encoding metadata of locations $[0,...,y-1]$ in the pre-fetched row in Cycle 1 is summed up through a binary adder tree
and then added with the start address of this row to obtain the address of the target non-zero element in the non-zero element array; \underline{Cycle 3}: the target non-zero element is fetched using the address calculated in Cycle 2. 

\begin{figure}[b]
\vspace{-1.5em}
  \centering
  \includegraphics[width=0.82\linewidth]{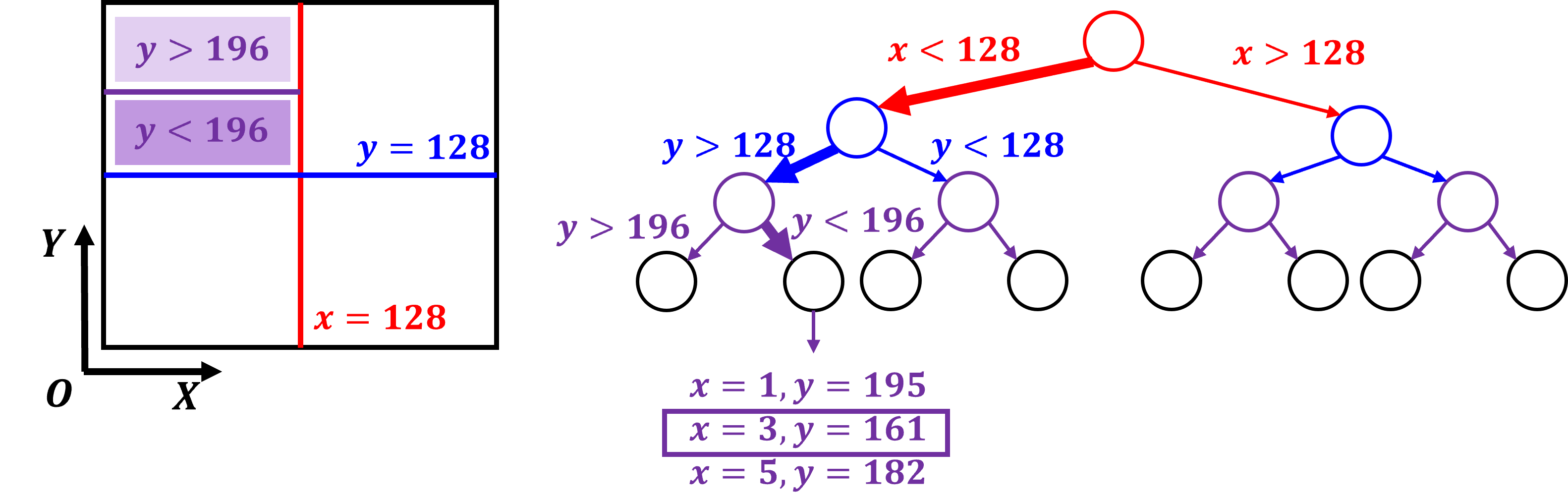}
\vspace{-1.3em}
\caption{An example of decoding the element at location ($x$=3, $y$=161) in the adopted coordinate-based decoding \& searching using a binary search tree.}
\vspace{-0.5em}
\label{fig:CoordEnoding}
\end{figure}

\begin{figure}[t]
  \centering
\includegraphics[width=0.95\linewidth]{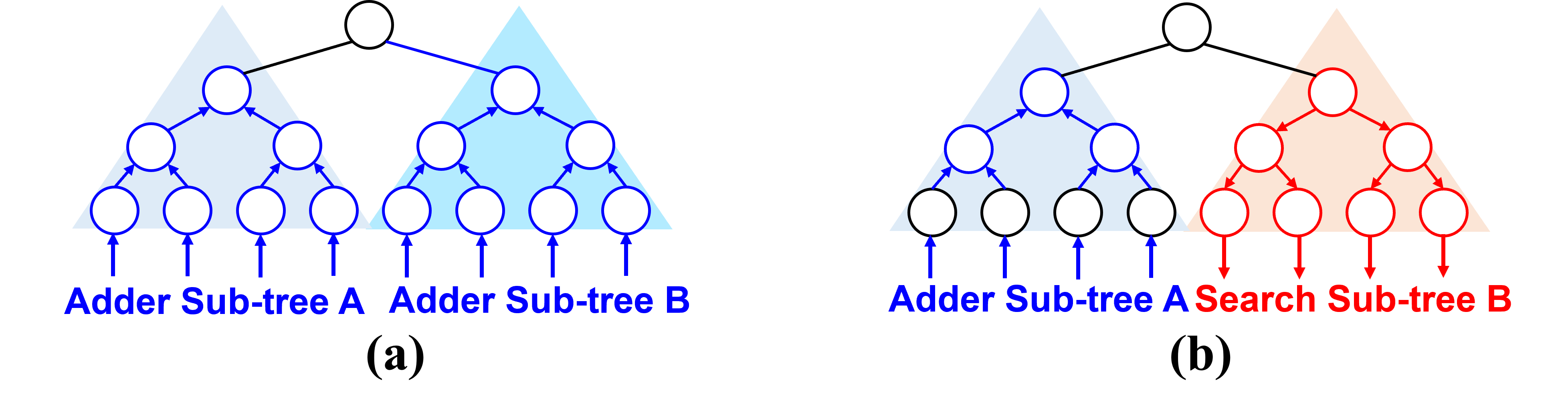}
\vspace{-1.5em}
\caption{Different reconfigurable modes of the dual-purpose bi-direction adder \& search tree unit.}
\vspace{-1.5em}
\label{fig:DBAQTree_Mode}
\end{figure}

\begin{figure}[b]
\vspace{-1em}
  \centering
\includegraphics[width=0.85\linewidth]{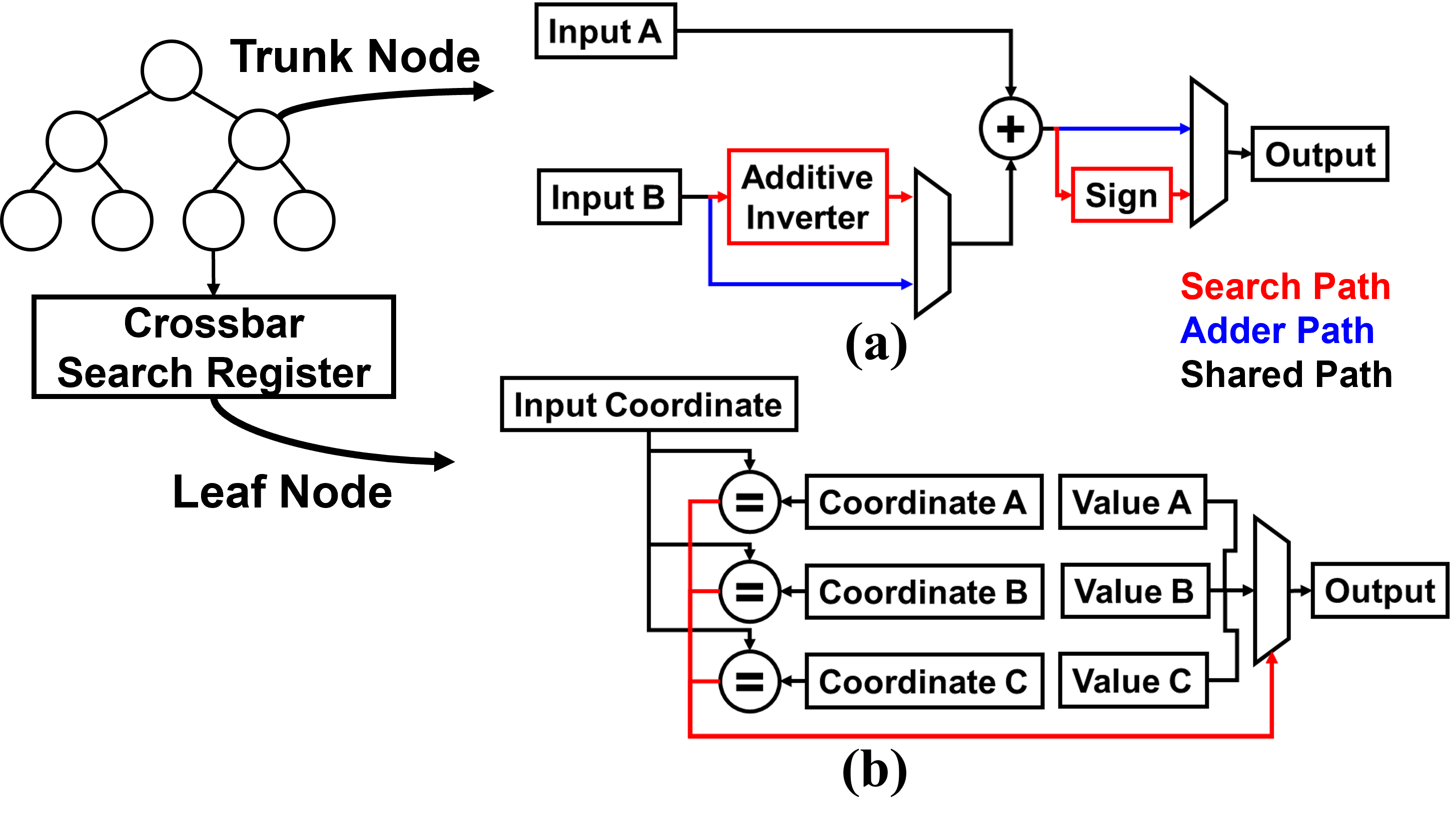}
\vspace{-2em}
\caption{The dual-purpose bi-direction adder \& search tree.}
\label{fig:DBAQTree_Reconf}
\end{figure}

\begin{table*}[!b]
\vspace{-0.5em}
\caption{A summary of the considered devices' specifications.}
\centering
\vspace{-1.3em}
  \resizebox{0.95\linewidth}{!}
  {
    \begin{tabular}{c||ccc||cccc}
    \toprule
    \multirow{3}{*}{\textbf{Device}}& \multicolumn{3}{c||}{\textbf{Edge Devices}} & \multicolumn{4}{c}{\textbf{Cloud Devices}} \\

    & \textbf{NVIDIA Jetson Nano~\cite{jetson_nano}} & \textbf{ICARUS~\cite{rao2022icarus}} &  \textbf{RT-NeRF-Edge} & \textbf{AMD Threadripper 3970x~\cite{ryzen}} & \textbf{NVIDIA RTX 2080Ti~\cite{2080ti}} & \textbf{NVIDIA  V100~\cite{v100}} & \textbf{RT-NeRF-Cloud}  \\
    
   \textbf{(Algorithm)}& \textbf{(TensorRF~\cite{chen2022tensorf})} &  \textbf{(Original NeRF~\cite{mildenhall2020nerf})} &  \textbf{(RT-NeRF)}&
     \textbf{(TensorRF~\cite{chen2022tensorf})}& \textbf{(TensorRF~\cite{chen2022tensorf})} & \textbf{(FastNeRF~\cite{garbin2021fastnerf})} & \textbf{(RT-NeRF)}  \\

    \midrule
    Computing & 1 (SM) &  N/A & 1 (Core) &32 (Core) & 68 (SM) &  80 (SM)& 30 (Core) \\
    Units & 128 (CUDA)  & N/A & 1 (SPU) + 1 (PPU)     &64 (Thread) & 4352 (CUDA) & 5120 (CUDA) & 30 (SPU) + 30 (PPU)  \\
    \midrule
    SRAM & 2 MB & 0.96 MB& 3.5 MB & 146 MB & 29.5 MB & 36 MB & 105 MB  \\
    \midrule
    Area (mm$^2$)& 118 & N/A & 18.85 & 296 & 754 & 815 & 565 \\
    \midrule
    Frequency & 0.9 GHz & 0.3GHz& 1 GHz& 3.7 GHz & 1.35 GHz & 1.5 GHz & 1 GHz  \\
    \midrule
    DRAM & LPDDR4-1600 &  N/A & LPDDR4-1600& DDR4-3200 & GDDR6 & HBM2 & HBM2 \\
    Bandwidth & 25.6 GB/s &  N/A & 17 GB/s& 95.4 GB/s& 616 GB/s & 900 GB/s & 510 GB/s \\
    \midrule
    Technology & 20nm & 40nm & 28nm& 7nm & 12nm & 12nm & 28nm  \\
    \midrule
    Typical Power & 10 W & 0.3 W & 8 W& 270 W & 250 W & 300 W & 240 W  \\
    \midrule
    Typical FPS & 0.01 & 0.03 & \textbf{45}& 0.5 & 0.8 & 200 & \textbf{1300}  \\
    \bottomrule
    \end{tabular}
    }
  \label{tab:platform}
\end{table*}

\textbf{Coordinate-Based Sparse Encoding Format and Dual-Purpose Bi-Direction Adder \& Search Tree.}
Inspired by~\cite{dave2021hardware}, we adopt a coordinate-based (i.e., COO) sparse encoding format for matrices/vectors with a high sparsity ($\geq$ 80\%).
In addition, we adopt a binary search tree based decoding method as illustrated in Fig.~\ref{fig:CoordEnoding} to better match the corresponding coordinate-based encoding format. 
Specifically, each node of the search tree will handle one part of the decoding process. For example, the lines in bold in Fig.~\ref{fig:CoordEnoding} indicate a search path for $x=3, y=161$, where each node is responsible for one comparison (e.g., $x<128$, $y>128$, $y<196$). In the leaf node, there is a one-to-one match crossbar that directly fetches the value of $x=3, y=161$.

The binary search tree can suffer from under-utilized in scenarios when most of the processed matrices' sparsity ratio is relatively low ($<$ 80\%) and bitmap-based encoding is used. For example, such scenarios are frequently observed since 32\%/68\% of the overall matrices and vectors are of high/low sparsity and utilize the coordinate-/bitmap-based encoding formats, respectively. 
To ensure a high hardware utilization, we combine the binary search tree for sparse decoding and the adder tree for computing the features (e.g., densities) of the pre-existing points (as demonstrated in Eq.~\ref{eq:tensorf_decomp}), and propose a dual-purpose bi-direction adder \& search tree unit.
Specifically, as illustrated in Fig.~\ref{fig:DBAQTree_Mode}, the proposed dual-purpose bi-direction adder \& search tree unit is designed to be reconfigured between (1) an adder tree with multiple adder sub-trees (Fig.~\ref{fig:DBAQTree_Mode} (a)) for the low sparsity scenario ($<$ 80\%) and (2) a mixed tree with adder sub-trees and search sub-trees (Fig.~\ref{fig:DBAQTree_Mode} (b)) for the high sparsity scenario ($\geq$ 80\%). 
Fig.~\ref{fig:DBAQTree_Reconf} shows the corresponding hardware implementation: Each trunk node can be reconfigured between an adder or a comparator by simply turning on/off an additive inverter before the adder or turning off/on a sign detector after the adder, respectively (Fig.~\ref{fig:DBAQTree_Reconf} (a)). Meanwhile, each leaf node in Fig.~\ref{fig:DBAQTree_Reconf} (b) is equipped with a crossbar-based search register to decode the target element when being configured as a search tree. 

%% file: sections/5-Experiments.tex
\section{Evaluation}
\subsection{Experiment Setup}

\textbf{Datasets \& Baselines.}
To evaluate the performance of the proposed RT-NeRF, we conduct experiments on the eight datasets of Synthetic-NeRF~\cite{mildenhall2020nerf}. For the evaluation baseline hardware devices, we adopt two categories of devices: edge and cloud devices, and consider a total of five baseline devices, including commercial GPUs and CPUs as well as a dedicated NeRF accelerator.
Specifically, for edge devices, we choose NVIDIA Jetson Nano~\cite{jetson_nano} (commonly-used edge GPU) and ICARUS~\cite{rao2022icarus} (a dedicated ASIC accelerator for NeRF); For cloud devices, we select NVIDIA V100~\cite{v100} (commonly-used cloud GPU with large GPU memory), NVIDIA RTX 2080Ti~\cite{2080ti} (commonly-used cloud GPU), and AMD Threadripper 3970X~\cite{ryzen} (common-used cloud CPU, 1 core is used in~\cite{chen2022tensorf}). Table.~\ref{tab:platform} summarizes the considered devices' specifications. 
 
\begin{figure*}[!b]
\vspace{-1em}
  \centering
\includegraphics[width=0.95\linewidth]{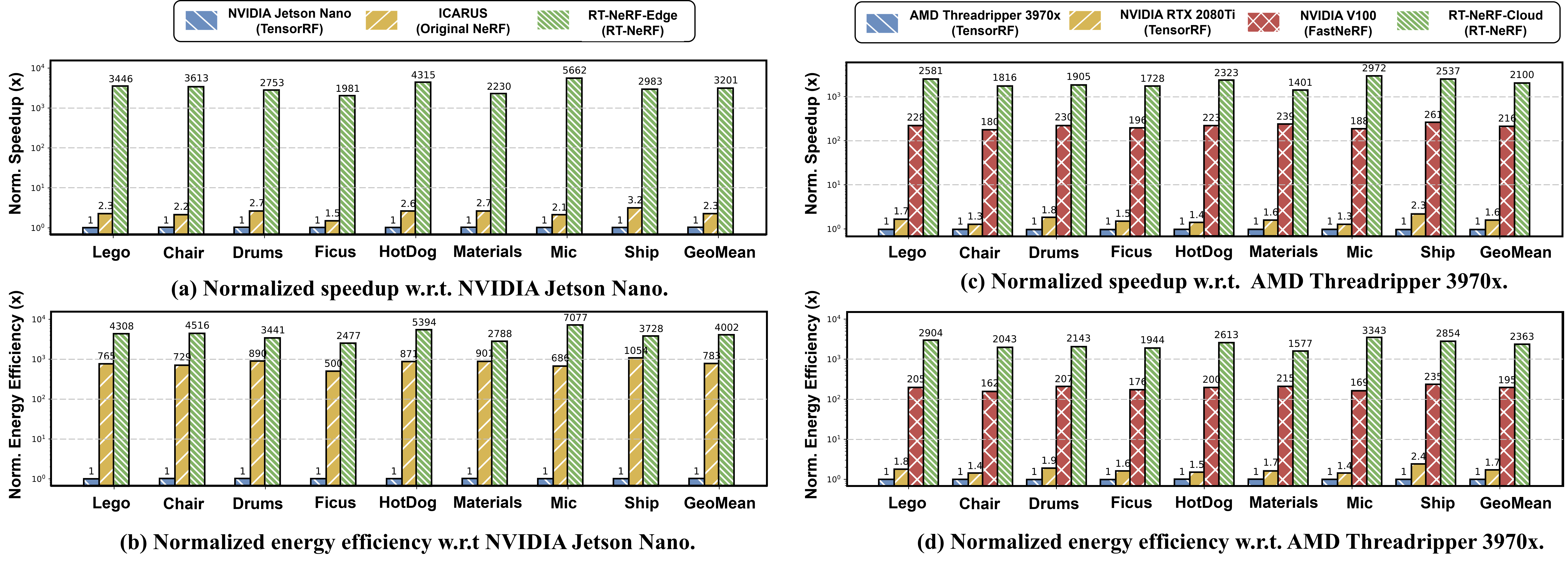}
\vspace{-1.5em}
\caption{The normalized speedup and energy efficiency achieved by our proposed RT-NeRF and five baseline devices on the eight datasets of Synthetic-NeRF~\cite{mildenhall2020nerf}. All the legends follow the ``device (algorithm)'' format.}
\vspace{-1em}
\label{fig:bench_chart}
\end{figure*}

\textbf{RT-NeRF Hardware Implementation.}
We implement the PPU in the proposed RT-NeRF accelerator with Verilog, and then synthesize, place \& route the design based on a commercial 28nm CMOS technology using Cadence Genus \& Innovus~\cite{cadence}. Additionally, we develop a cycle-accurate simulator to simulate the performance of our PPU, for which the unit computation cost is derived from the post-layout simulation. After that, we verified the simulator against the Verilog implementation to ensure its correctness. 
The SPU in our RT-NeRF accelerator is simulated using a modified simulator based on SonicBOOM~\cite{zhaosonicboom} which is an open source out-of-order RISCV core.
For a fair comparison with both the edge and cloud baseline devices, we configure two RT-NeRF hardware settings accordingly: one corresponds to the edge device (denoted as RT-NeRF-Edge) and the other corresponds to the cloud device (denoted as RT-NeRF-Cloud). Specifically,
for RT-NeRF-edge, we set the corresponding hardware configuration such that RT-NeRF-edge can achieve > 30FPS throughput requirement for all eight datasets of Synthetic-NeRF~\cite{mildenhall2020nerf}, which results in an average of 45FPS over the eight datasets. For RT-NeRF-Cloud, we configure the hardware resources to match the power of a RTX 2080Ti~\cite{2080ti} GPU. The total area of our RT-NeRF-Edge and RT-NeRF-Cloud accelerators are 18.85$mm^2$ and 565$mm^2$, respectively. 
The detailed hardware configurations are summarized in Tab.~\ref{tab:platform}.

\subsection{Algorithm Evaluation}
To evaluate the effectiveness of our RT-NeRF algorithm in Sec.~\ref{sec:algorithm}, in addition to the changes of the runtime breakdown before and after applying our proposed algorithm on 2080 Ti~\cite{2080ti} in Fig.~\ref{fig:technique}, we benchmark it with SOTA efficient NeRF algorithms in terms of achieved PSNR~\cite{hore2010image} (a higher value indicates better rendering quality) on eight datasets of Synthetic-NeRF~\cite{mildenhall2020nerf} in Tab.~\ref{tab:algs_benchmark}. We can see that (1) our RT-NeRF algorithm surpasses all the SOTA efficient NeRF algorithms except TensoRF~\cite{chen2022tensorf} in terms of the rendering quality (i.e., $\uparrow$0.78 $\sim$ $\uparrow$1.89 PSNR, averaged over the eight datasets), while achieving a higher energy efficiency (i.e., $\uparrow$5.1$\times$ $\sim$ $\uparrow$4002$\times$, averaged over the eight datasets) as suggested in Fig.~\ref{fig:bench_chart}; (2) Compared with TensoRF~\cite{chen2022tensorf}, the average PSNR achieved by our proposed algorithm is only $\downarrow$0.21 than TensoRF~\cite{chen2022tensorf}, which is caused by approximating the cubes as a ball in Step {\color{black}{\ding{183}}}-{\color{black}{\ding{172}}}-a, however, our RT-NeRF algorithm can reduce the rendering latency by $\downarrow$1.4$\times$ over TensoRF~\cite{chen2022tensorf} on commercial devices as suggested in Fig.~\ref{fig:technique}.

\begin{table}[t]
\caption{Benchmark our proposed RT-NeRF with SOTA efficient NeRF algorithms in terms of the PSNR~\cite{hore2010image} (higher value represents better rendering quality).}
\centering
\vspace{-1em}
  \resizebox{0.95\linewidth}{!}
  {
    \begin{tabular}{c||c||cccccccc}
    \toprule
    Methods  & Avg. & Chair & Drums & Ficus & Hotdog & Lego & Materials & Mic & Ship \\
    \midrule
    NeRF~\cite{mildenhall2020nerf} &  31.01 & 33.00 & 25.01 & 30.13 & 36.18 & 32.54 & 29.62 & 32.91 & 28.65 \\
    ICARUS~\cite{rao2022icarus} & 30.21 & 33.14 & 30.38 & 28.57 & - & 29.48 & - & - & 29.48 \\
    FastNeRF~\cite{garbin2021fastnerf} & 29.90 & 32.32 & 23.74 & 27.79 & 34.72 & 32.27 & 28.88 & 31.76 & 27.68 \\
    TensoRF~\cite{chen2022tensorf} & 32.00 & 34.68 & 25.37 & 32.30 & 36.30 & 35.42 & 29.30 & 33.21 & 29.46 \\
    \midrule
    \textbf{RT-NeRF (Ours)} & 31.79 & 34.52 & 25.05 & 32.11 & 36.02 & 35.21 & 29.10 & 33.01  & 29.27 \\
    \bottomrule
    \end{tabular}
    }
  \label{tab:algs_benchmark}
\vspace{-1em}
\end{table}

\subsection{Hardware Evaluation}
\label{sec:hardware_eval}

Fig.~\ref{fig:bench_chart} presents the efficiency improvements achieved by the proposed RT-NeRF in comparison with the five baselines on the eight datasets of Synthetic-NeRF~\cite{mildenhall2020nerf}.  
Compared with the edge devices, the proposed RT-NeRF on average offers 3201$\times$, 1391$\times$ speedup and 4002$\times$, 5.1$\times$ energy efficiency over Jetson Nano~\cite{jetson_nano} and ICARUS~\cite{rao2022icarus} (Fig.~\ref{fig:bench_chart} (a) and (b)), respectively. 
Compared with cloud devices, the proposed RT-NeRF on average offers 2100$\times$, 1312$\times$, and 9.7$\times$ speedup and 2363$\times$, 1390$\times$, and 12.1$\times$ energy efficiency over AMD Threadripper 3970X~\cite{ryzen}, NVIDIA 2080Ti~\cite{2080ti}, and NVIDIA V100~\cite{v100}  (Fig.~\ref{fig:bench_chart} (c) and (d)), respectively. 
It is worth noting that when running the same proposed RT-NeRF algorithm on both NVIDIA 2080Ti~\cite{2080ti} and our proposed accelerator, as visualized in Fig.~\ref{fig:technique} (b) and (c), respectively, our RT-NeRF-Edge accelerator can further decrease 97\% of the latency in Step {\color{black}{\ding{183}}}-{\color{black}{\ding{173}}} and achieves 24$\times$ speed up over NVIDIA RTX 2080Ti~\cite{2080ti}, validating the effectiveness of our proposed accelerator. 

%% file: sections/6-Related_Works.tex
\section{Related Works}
A concurrent work of our proposed RT-NeRF is ICARUS~\cite{rao2022icarus}, which proposes an architecture for the vanilla MLP-dominated NeRF-based rendering process. In contrast, to the best of our knowledge, our proposed RT-NeRF is the first algorithm-hardware co-design acceleration of NeRF. Furthermore, our proposed RT-NeRF is built on top of SOTA efficient NeRF algorithms, and dedicated to accelerate their unique bottleneck operations other than merely the MLP inference (see Sec.~\ref{sec:background_motivations}), thus achieving 1393$\times$ speedup and 5.1$\times$ energy efficiency over ICARUS~\cite{rao2022icarus} as shown in Sec.~\ref{sec:hardware_eval}.

%% file: sections/7-Conclusion.tex
\section{Conclusion}
We propose, develop, and validate RT-NeRF, the first algorithm-hardware co-design acceleration of NeRF. 
On the algorithm level, RT-NeRF integrates an efficient rendering pipeline for leveraging the sparsity of pre-existing points and a coarse-grained view-dependent rendering ordering to avoid processing invisible points. On the hardware level, RT-NeRF adopts a hybrid encoding scheme and integrates both a dual-purpose bi-direction adder\&search tree unit and a high-density sparse search unit for ensuring efficient sparse decoding. 
We believe our work can open up an exciting perspective towards real-time NeRF solutions.